# Spontaneous Symmetry Breaking in Metal Adsorbed Graphene Sheets

A. F. Jalbout

*Departamento de Investigacion en Fisica (DIFUS), Universidad de Sonora, Hermosillo, Sonora, C.P. 83210*

T.H. Seligman

*Instituto de Ciencias Físicas, Universidad Nacional Autónoma de México, Cuernavaca, México*

*Centro Internacional de Ciencias, Cuernavaca, México*

Graphene has received a great deal of attention and this has more recently extended to boron nitride sheets (BNS) with a similar structure. Both have hexagonal lattices and it is only the alternation of atoms in boron nitride, which changes the symmetry structure. This difference can for example be seen in the mean field equations, which for the corners of the Brillouin Zone are Dirac equations[1]. For the case of graphene (equal atoms) we have the equation for massless particles, while for Boron Nitride has a finite gap and is more near a Dirac equation with mass near this gap.. Carbon structures in general and in particular also graphene can adsorb electron donors, such as alkaline atoms or molecules with a dipole moment. Typically these atoms and the dipoles can only attach in the sense to donate electron density[2-5]. Some results for small sheet like structures are available[6]

In the present paper we analyze the adsorption of Lithium atoms to graphene and BNS. Lithium atoms in this case are paradigmatic, in that other alkaline atoms behave similarly. Calculations are carried out with up to two sites of such sheets, adsorbing one and two Lithium atoms. In the case of two atoms these can be on the same side or on opposite sides of the sheet. The interest in such structures is multiple.

In previous studies we demonstrated that in the $Li^+$ case, a charge of ~0.8 is transferred to the surface of the $C_{60}$ species[2]. Modification of the surface electron configuration was shown to yield increased reactivity. Based on these and similar results for nanotubes, we shall explore adsorption of alkaline atoms, particularly lithium to graphene and to boron nitride sheets.

The basic results of our previous calculations suggests that in fullerenes the $Li@C_{60}$ (whereby Li is encapsulated in the $C_{60}$ frame) system can roughly be viewed as $Li^+@C_{60}^-$ since the electronic density of the Li atom primarily is transferred to the fullerene surface[2]. Analysis of the charge population showed that the majority of this density is concentrated in a distinct region of the surface. In the present analysis we will discuss how charge mediated bond stretching caused by metal adsorption can influence the chemical and physical properties of carbon and boron nitride sheets. Among these properties we shall find, as the central result of this letter, spontaneous symmetry breaking. This effect will be explained both in terms of a qualitative discussion of bond stretching and as a generalized Jahn Teller effect[7].

We must recall, that the sheets under discussion have an in plane symmetry of translations and rotations and reflections as well as a mirror symmetry at the plane of the sheet. Unsurprisingly our calculations reveal a deformation of the lattice which violates this mirror symmetry if we adsorb a single atom. If two atoms are adsorbed they can either be fixed to independent sites on the same or opposite side of the sheet or the second atom can find a stable position on the other side of the sheet right opposite to the first one. The surprise is, that the deformation in the latter case is larger than in the former. For one adsorbed atom any Born-Oppenheimer Hamiltonian will violate the mirror symmetry, but if two atoms are adsorbed on opposite sides of the sheet a symmetric configuration exists. Surprisingly this symmetric nuclear configuration is unstable in any of the approximations we used, i.e. in Hartree-Fock calculations with and without electron correlations as well as in density functional computations. For graphene we shall show this effect for increasing sizes of lattices from 6 to 20 rings, to make sure that we do not see finite size effects. These surface sizes have been shown to be adequate in the calculation of charge transfer in realistic systems [3-5]. The symmetry breaking is present throughout this range.

Furthermore sample BN sheets were used to confirm that spontaneous symmetry breaking is also seen in this structure of lesser symmetry. In this case we limited our calculations to sheets of nine rings and found very similar results to the corresponding Carbon sheet. This is particularly important as one might suspect, that the gapless nature of the spectrum in the mean field approximation for graphene[1] might be responsible for the symmetry breaking. Yet for BN sheets the mean field equation is a Dirac equation with finite mass, and thus with a gap in the spectrum. Thus the possibility, that the gapless spectrum is responsible for the symmetry breaking is disproven.

We conclude, that the symmetry breaking may be related to the Jahn-Teller effect[7]. We recall that this effect was originally proven for molecules and indeed for each point group individually, but there is a little known general proof by Ruch and Schoenhofer[8], based on general symmetry concepts. We shall see that the present instability of the symmetric configuration can be explained along these lines. This indicates that the physics of the Jahn Teller effect may reach beyond the common field of applications and shows that perturbations in sheets will behave basically differently than in bulk solids. Yet we must bear in mind, that a quantum phase transiton may instead be responsible for the spontaneous symmetry breaking.

We shall first present the numerical results illustrated for different sheets, such as to clarify the phenomenology of the symmetry breaking we observe. Next we outline the basic idea of the Ruch and Schoenhofer argument[8] and discuss how it applies to the present configuration. Finally we give a simple physical argument, why this should occur, based on the specific structure we analyze, and we try to draw conclusions for other systems.

All of the quantum chemical computations were performed with the GAUSSIAN03[9] program codes. Since the systems are relatively large the geometry optimizations were performed with the Hartree-Fock (HF) method[10-12]. Higher level calculations (DFT and Electron correlations over the Hartree Fock results) were performed for specific geometries to confirm consistency of the mean field results. These tests were important, because experiments on the systems explored are not yet available.

To do the calculations we have selected for tis work the STO-3G* basis which is an acronym for Slater-Type-Orbitals simulated by three Gaussians added together[13-16]. In this basis set the coefficients of the Gaussian functions are adjusted to yield reasonable fits to the Slater orbitals. Some of the smaller sheets were calculated with larger basis sets to ensure that the results are reliable. To verify the mean field effects with the HF equations the calculations were complemented by test density functional theory (DFT) computations. The B3LYP method[13-16] coupled to the same basis sets was used for this purpose. This method is the Becke Three Parameter Hybrid Functionals which are composed of $A*E_X^{Slater}+(1-A)*E_X^{HF}+B*\Delta E_X^{Becke}+E_C^{VWN}+C*\Delta E_C^{non-local}$ by which A, B, and C are the constants determined by Becke and fitting to the G1 molecule set. Furthermore we have performed calculations and tested our results with sodium and potassium atoms and they were consistent with those for lithium. They will not be reported on for space reasons.

In our case the situation is slightly different: The presence of the adsorbed atoms already violates the in plane symmetry, but as they are not really bound we might consider this as a perturbation. Yet the donated electrons form an integral part of the electron structure of the lattice. The localization of the donated electrons breaks the in plane symmetry of the electron distribution, producing conditions that the proof of Schoenhuber and Ruch may be valid. Yet it is difficult in such a many body problem to assert or negate the degeneracy.

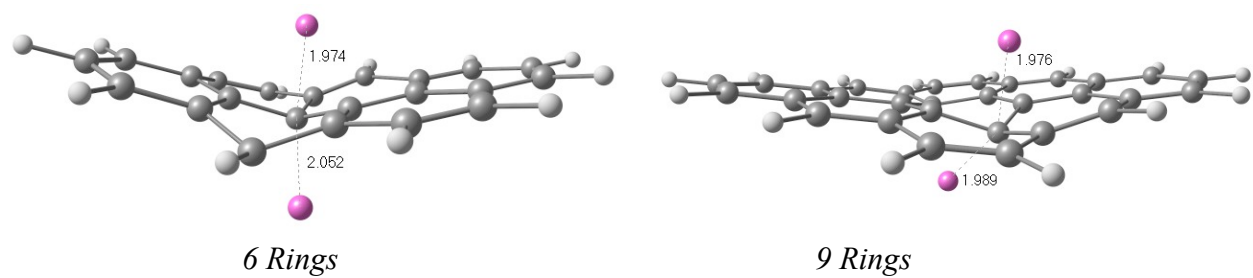

*6 Rings*  *9 Rings*

**Figure 1.** Double Li atom adsorption to two carbon sheets.

For the six ring case we can see that the energies of interaction are about 36.4 kcal/mol which has an intermolecular separation of around 2 Å (from Figure 1). The absorption energy of the Li atom to the first species (which is the 6 ring system) is around 59.1 kcal/mol with an intermolecular separation of 2.4 Å. At the HF/STO-3G level of theory the Mulliken partial charge on Li is around 0.26e. It is quite interesting to note that we can strongly observe the symmetry breaking in this species which is consistently observed in these sheets. In both the HF and DFT cases we observe this unusual effect.

We have performed a test HF/6-311G** natural orbital population analysis (NOPA) which yields a partial charge on Li of 0.85e with the excess electron (transferred from the Li atom to the surface) occupying a localized region on the hydrocarbon surface. While NOPA charges are more accurate it is our intention to discuss general trends upon metal absorption to an extended planar aromatic molecular surface. Calculations of the Mulliken partial charges suggest that the Li atoms equally donate around 0.8e to the molecular surface independent of the surface employed.

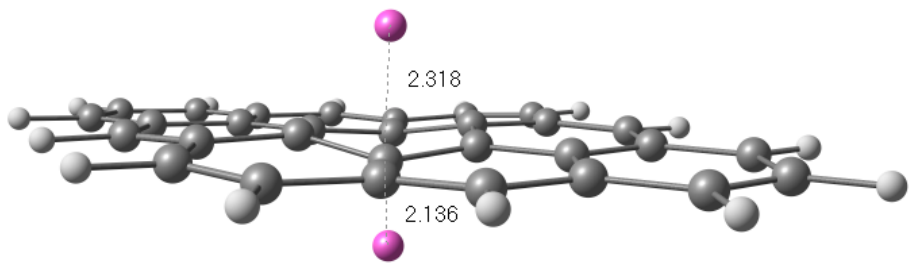

**Figure 2.** The DFT-B3LYP/3-21G* result of double Li adsorption to the nine-ring graphene carbon sheet.

A further increase of the system to 9 rings creates an intermolecular separation of 2.4 Å. The reason we have not included the values of all the systems is in interest of space. It is our observation that the Li atoms tend to form an attractive polarization force when they are on the molecular surface. Even at larger distances there is an interatomic interaction, because when they are segregated on extended molecular surfaces they favor small separations. It is also important to note that the results are consistent within density functional theory (DFT) which removes problems with mean field effects. The results within DFT show the same symmetry breaking as seen in Figure 2. In both cases we can easily see how the symmetry breaking plays a significant role.

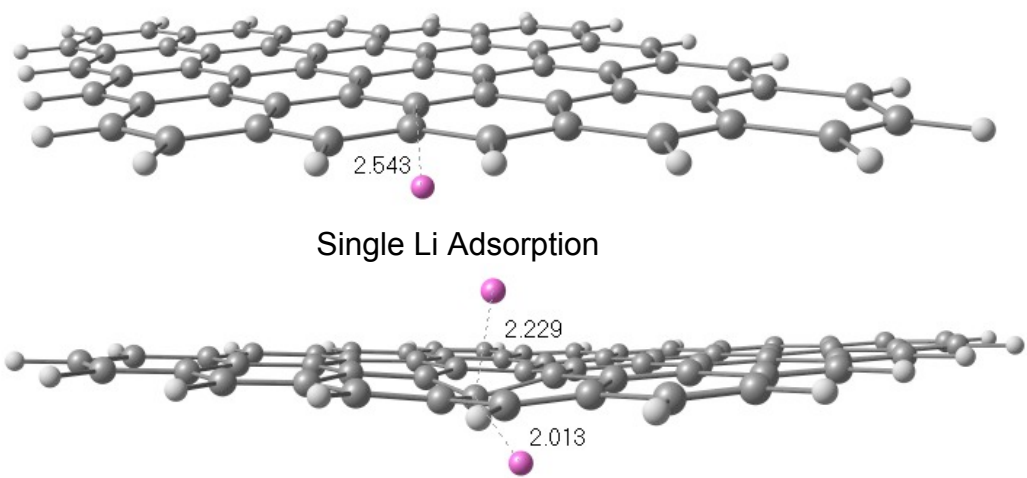

Single Li Adsorption

Double Li Adsorption

**Figure 3.** Li adsorption to a 20 graphene ring systems.

Another important question that arises is the local nature of the symmetry breaking. To further demonstrate the efficacy of the proposal that we present in Figure 3 we present the results of Li atom adsorption to a 20 ring system as we can see the two atoms also cause distortion of symmetry upon attachment to the surface. The DFT result shows how the extended surface exhibits local symmetry distortions upon metal bonding to the sheet. Even the single Li adsorption case causes minor distortions along the sheet, but it is more prolific when two Li atoms bind to a point. The exact deviation between adsorption energies is shown in Table 1.

| No. Rings | Single | Double |
|---|---|---|
| 6 | 59.07 | 11.13 |
| 7 | 99.30 | 21.23 |
| 8 | 103.72 | 16.32 |
| 9 | 109.28 | 22.23 |
| 17 | 287.71 | 56.88 |
| 18 | 275.84 | 62.23 |
| 19 | 279.43 | 81.77 |
| 20 | 290.56 | 90.07 |

**Table 1.** Energies (kcal mol$^{-1}$) and distances (angstroem) of single and double (to the same site) Li adsorbtion to the carbon sheets

A simpler understanding is possible and indeed rather obvious. The adsorbed charge lengthens the bonds the region of localized enhanced electron density. While for a small sheet a deformation in the plane would be possible in a large or infinite sheet this cannot happen, and the plane will have to buckle to either side. The sequence in which the two atoms are adsorbed will obviously determine the side to which the longer bonds will push the site. This is analogue to the explanation found for the Jahn Teller effect in point symmetries and thus rounds the picture we have. Both the effect and its understanding are thus intimately related to the Jahn Teller effect and it becomes plausible why graphene and BNS behave in the same way. Indeed we would expect any flat 2-D Lattice
to behave in a similar way under any bond lengthening perturbation.
 This would for example include adsorption of a dipole or a defect which replaces an atom by one bonded with longer bonds. A more challenging question arises, as to what would happen if we introduce an effect with shortened bonds. The second explanation would predict just that- a defect in the plane, but the general argument will not allow that. It still requires symmetry breaking. This must then imply a mirror symmetry breaking deformation which adjusts nearby distances and a healing by bond stretching on a larger scale. This would likely be quite unfavorable energetically and thus would not occur under adsorption but would have to happen with some built in defect such as substitution of another atom that has shorter bonds to the neighbors.

Summarizing we have shown that adsorption of electron donor atoms on graphene and Boron nitride surfaces is possible, and that it has a local effect on the electron density. Similar behavior for adsorption for dipoles can be inferred from similar calculations on nano tubes and nanocones[14] and corresponding calculations are under way shortly. This opens the way to applications mentioned above.

The reason that these molecular surfaces were selected corresponds to the fact that the extensive aromatic nature of the benzene rings permits an adequate adaption of the excess electrons of Li to be transferred. In other words, the Li atom contributes to the electronic density on localized regions of the molecular surface.

The calculations effectively reveal that charge transfer effects occur between the metal atom and the molecular surface. The absorption properties when the two Li atoms are separated (either on the same side or opposing) are favored on the molecular sheets. Another very interesting concept is that symmetry breaking at the Li-molecular surface interface is observed in certain systems. This effect is more prolific in smaller systems whereby the Li atoms tend to occupy similar regions on the molecular surface. Such behavior cannot be observed in reduced molecular frameworks since the Li atoms will tend to aggregate and dimerize.

The fact that charge is transferred to the systems considered means that the molecular surfaces can aid in the breaking apart of bound systems. The surfaces can in fact lower the dissociation energy of metal dimers leading to the storage of excess electronic density in regions along the molecule. More interestingly, the concepts used are intrinsic to the metal adsorption without being affected by the type of sheet being used.

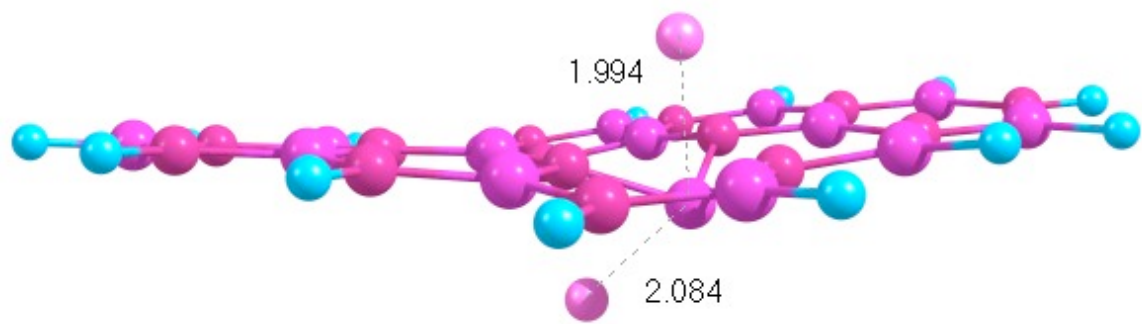

**Figure 4.** Double Li adsorption to a 9 membered BN ring system that clearly shows the spontaneous symmetry breaking.

Finally, in Figure 4 we present the results of double Li binding to a nine ring BN sheet computed at the DFT-B3LYP level of theory. Interestingly, we see the same results as before with a similar intermolecular separation in the cases calculated. The single Li adsorption energies for the BN sheet are 13.8 and 15.92 kcal/mol at the HF and DFT levels of theory, respectively. The double adsorption energies for Li are 9.8 (HF) and 37.3 (DFT) kcal/mol. The adsorption energies are higher suggesting that there is a higher degree of symmetry breaking in these cases as compared to the carbon graphene sheets.

We have further shown that adsorption may occur on both sides of the same site of the sheet, but that the resulting configuration is not symmetric under reflection in the plane due to a behavior closely related to the Jahn Teller effect. This relation allows us also to predict similar symmetry breakings with other kinds of local perturbations that break the translational symmetry and in principal allow mirror symmetric Born Oppenheimer configurations. An interesting question may close the view of this problem. Assume we could have a flat 2-D quasicrystal and we can adsorb Lithium in the same way. Would spontaneous breaking of the reflection symmetry occur? The question sounds somewhat artificial at this point, but it would be challenging to see whether the Schoenhofer-Ruch argument extends to this situation or not. In this work we not only attempt to set-up the ground work for future investigations but question basic nature that can lead to potentially new avenues of research.

Acknowledgements

We appreciate support from the CONACyT project 79613 and one of us (A.J.) the hospitality of CIC on various occasions.